\documentclass[conference]{IEEEtran}
\IEEEoverridecommandlockouts
\usepackage{cite}
\usepackage{amsmath,amssymb,amsfonts}

\usepackage{algorithm}
\usepackage{algpseudocode} 
\usepackage{graphicx}
\usepackage{textcomp}
\usepackage{xcolor}
\usepackage{float}
\usepackage{graphicx}
\usepackage{subcaption}

\def\BibTeX{{\rm B\kern-.05em{\sc i\kern-.025em b}\kern-.08em
    T\kern-.1667em\lower.7ex\hbox{E}\kern-.125emX}}
\begin{document}

\title{Deep Learning-Enabled System Diagnosis in Microgrids: A Feature-Feedback GAN Approach
\\

\thanks{This work was supported by Korea Institute of Energy Technology Evaluation and
Planning (KETEP) grant funded by the Korea government (MOTIE) (RS-2023-00231702, Development of an open-integrated platform for distributed renewable energy systems)}
}

\author{\IEEEauthorblockN{1\textsuperscript{st} Swetha Rani Kasimalla}
\IEEEauthorblockA{\textit{Electrical
and Computer Engineering} \\
\textit{University of Michigan}\\
Dearborn, USA \\
sweraka@umich.edu}
\and
\IEEEauthorblockN{2\textsuperscript{nd} Kuchan Park}
\IEEEauthorblockA{\textit{Electrical
and Computer Engineering} \\
\textit{University of Michigan}\\
Dearborn, USA \\
kuchan@umich.edu}
\and
\IEEEauthorblockN{3\textsuperscript{rd} Junho Hong}
\IEEEauthorblockA{\textit{Electrical
and Computer Engineering} \\
\textit{University of Michigan}\\
Dearborn, USA \\
jhwr@umich.edu}
\and
\IEEEauthorblockN{4\textsuperscript{th}  Young-Jin Kim}
\IEEEauthorblockA{\textit{Electrical Engineering} \\
\textit{Pohang
University of Science and Technology}\\
Pohang, South Korea \\
powersys@postech.ac.kr
}
\and
\IEEEauthorblockN{5\textsuperscript{th} HyoJong Lee}
\IEEEauthorblockA{\textit{Principal Specialist Engineer} \\
\textit{DTE Energy}\\
Detroit, USA \\
hyojong.lee@dteenergy.com}

}

\maketitle

\begin{abstract}
The increasing integration of inverter-based resources (IBRs) and communication networks has brought both modernization and new vulnerabilities to the power system infrastructure. These vulnerabilities expose the system to internal faults and cyber threats, particularly False Data Injection (FDI) attacks, which can closely mimic real fault scenarios. Hence, this work presents a two-stage fault and cyberattack detection framework tailored for inverter-based microgrids. Stage 1 introduces an unsupervised learning model— Feature-Feedback Generative Adversarial Network (F2GAN)—to distinguish between genuine internal faults and cyber-induced anomalies in microgrids. Compared to conventional GAN architectures, F2GAN demonstrates improved system diagnosis and greater adaptability to zero-day attacks through its feature-feedback mechanism. In Stage 2, supervised machine learning techniques, including Support Vector Machines (SVM), k-Nearest Neighbors (KNN), Decision Trees (DT), and Artificial Neural Networks (ANN) are applied to localize and classify faults within inverter switches, distinguishing between single-switch and multi-switch faults. The proposed framework is validated on a simulated microgrid environment, illustrating robust performance in detecting and classifying both physical and cyber-related disturbances in power electronic-dominated systems.

\end{abstract}


\section{Introduction}
Today’s power systems are rapidly transforming with an emphasis on sustainability, operational efficiency, and resilience — much of which is driven by the emergence of microgrids and the integration of IBRs. IBRs significantly enhance grid flexibility, support high renewable penetration, and reduce power losses. However, their increasing reliance on digital control, real-time communication, and smart grid architectures introduces critical challenges, particularly concerning system reliability, privacy, and cybersecurity. Among the most prominent concerns are line faults and inverter-level faults. These physical faults and the system’s growing exposure to digital threats demand more advanced detection and mitigation mechanisms.

\subsection{Related Work}
Previous works such as~\cite{8590418} and~\cite{8329516} have explored open-circuit fault diagnosis in motor-driven systems, while~\cite{9544141} was the first to investigate open-circuit fault detection in a 7-level hybrid active neutral point clamped (7L-ANPC) multilevel inverter. Building on these foundations, more recent studies~\cite{10214285,10215460,10451077} have advanced the field by employing data-driven techniques for diagnosing open-switch faults. In parallel,~\cite{9970321} proposed an LSTM-based approach for identifying anomalies and physical faults, and~\cite{8731755,10689021} demonstrated how GANs can be effectively utilized to mitigate FDI attacks. However, limited research has been conducted on integrating these approaches to simultaneously evaluate both physical faults and FDI attacks within the same framework using the GAN model.

\subsection{Contributions}

This paper proposes a novel two-stage framework that integrates unsupervised anomaly detection with supervised fault classification to address the dual challenges of internal fault detection and FDI attack mitigation. In the first stage, a F2GAN is introduced. The F2GAN architecture consists of two key modules: a generator trained using feature matching loss to accurately replicate realistic fault patterns, and a discriminator optimized through feedback-driven training to distinguish internal faults from cyber-induced anomalies, even without requiring labeled attack data. The proposed F2GAN is benchmarked against a conventional GAN to validate its robustness, particularly under zero-day attack scenarios.

In the second stage, once a fault is detected, supervised learning models—SVM, KNN, DT, and ANN—are employed for fault localization and classification. This hybrid framework ensures resilient system diagnosis while enabling accurate classification of inverter faults, facilitating timely operational response.
Therefore, the key contributions of this work are fourfold. 
\begin{itemize}
   \item A novel unsupervised F2GAN architecture is developed, integrating feature matching to improve fault pattern learning and detection capability.
   \item The model is benchmarked against a conventional GAN, with both statistical and visual evaluation confirming its superior robustness in system diagnosis.
   \item The framework supports fault classification by distinguishing between different types of inverter faults using supervised machine learning.
  \item Multiple classifiers are evaluated to ensure the reliability and generalization of the proposed method for real-world microgrid scenarios.
\end{itemize}

\subsection{Paper Outline}
The remainder of this paper is structured as follows. Sections II and III describe the architecture of the microgrid and the mathematical formulation of the proposed two-stage framework. Section IV presents simulation results and performance analysis. Section V concludes the paper and outlines potential future work.

\section{Microgrid Framework and Threat Model}
\subsection{Microgrid framework and Vulnerability Zones}
\begin{figure}[!t]
\centering
\includegraphics[width=\columnwidth,keepaspectratio]{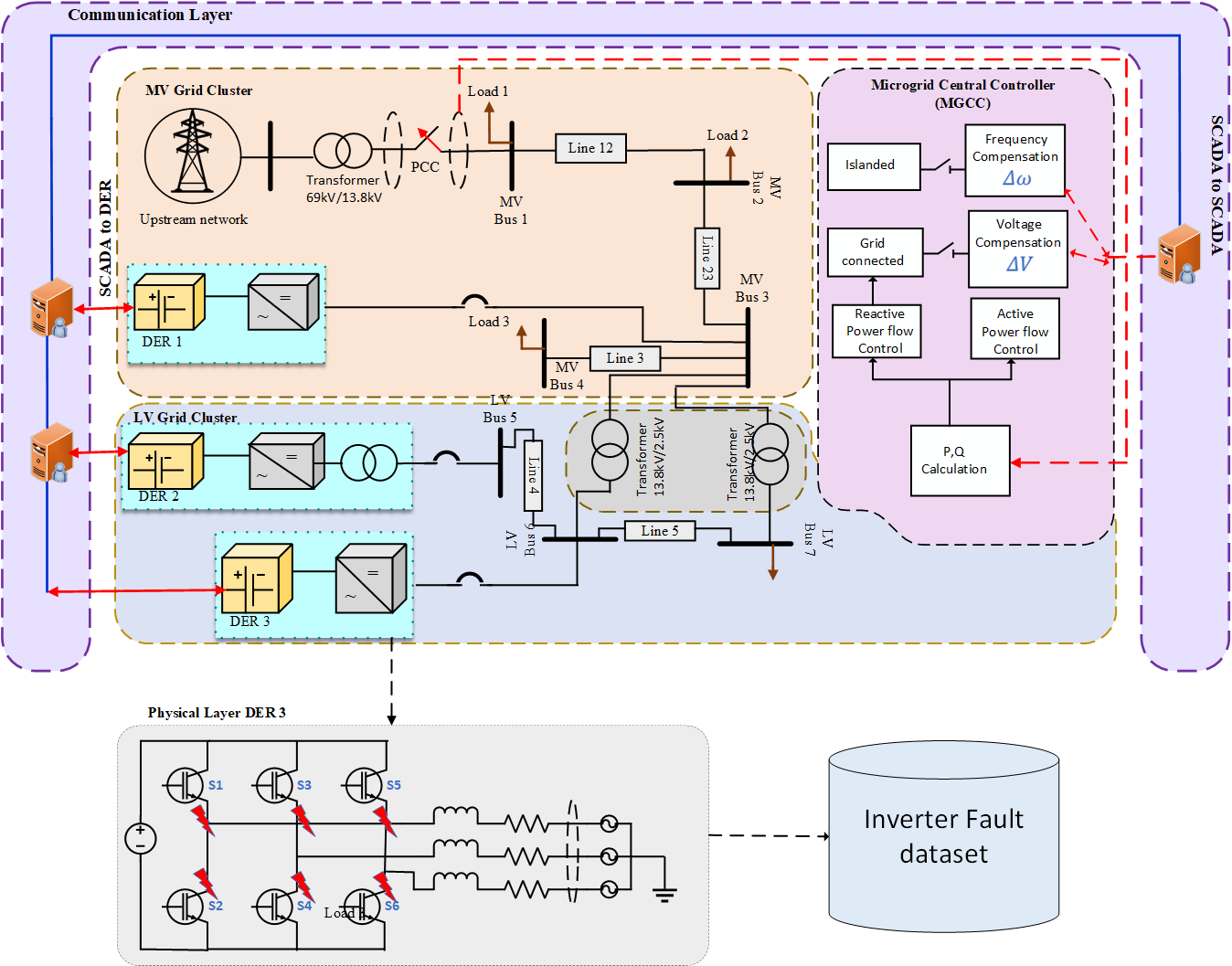}
\caption{Multi-layer microgrid framework illustrating DER integration used to simulate inverter faults.
The resulting data is collected for deep learning-based fault and anomaly analysis.}
\label{fig:MG}
\end{figure}

Fig. \ref{fig:MG} illustrates the multilayer microgrid framework utilized in this study. DER 1 is a battery-controlled inverter-based resource, DER 2 is a solar-based IBR, and DER 3 represents the battery controlled grid-connected system used to simulate multiple inverter fault scenarios for dataset collection. The Microgrid Central Controller (MGCC) coordinates transitions between grid-connected and islanded modes during fault conditions. The following layers describe the functional roles and vulnerabilities of each component, highlighting potential exposure to cyber-physical threats. The framework comprises four key layers that coordinate sensing, control, and power delivery, each playing a critical role in the operation and security of the system.

\subsubsection{Communication Layer}
The communication layer serves as the backbone for data exchange between the supervisory and physical layers. It transmits real-time current and voltage measurements from sensors to controllers across the grid. While essential for stable operation, this layer is highly susceptible to FDI attacks. Malicious intrusions at this layer can compromise measurement integrity, leading to corrupted decision-making and system-wide instability.
\subsubsection{Microgrid Central Controller Layer}
Acting as the brain of the system, the MGCC evaluates voltage and frequency data to determine the operational mode (grid-connected or islanded). It issues active/reactive power setpoints and voltage/frequency compensation commands to local DER controllers. This layer’s effectiveness directly depends on the accuracy of data received from the communication layer, making it vulnerable to any manipulated signals passed through FDI.
\subsubsection{Physical Layer (DERs and Local Controllers)}
This layer includes IBRs and their respective local controllers. It executes the operational setpoints provided by the MGCC, ensuring voltage and current control in real-time. However, its stability is directly affected by both cyber-injected measurement disruptions and physical inverter-level faults, making it a critical zone for fault detection and system protection.
\subsubsection{Inverter Fault Scenario}
This layer collects fault data from the system's physical layer under various operating and fault conditions to form the inverter fault dataset. As shown in Fig.~\ref{fig:MG}, the inverter, based on a three-leg topology, is subjected to all possible combinations of single-switch and multiple-switch faults. Scenarios are tested under dynamic conditions such as load changes, power fluctuations, and mode transitions (e.g., grid-connected to islanded mode). Voltage and current waveforms recorded at the inverter terminals serve as primary features for fault classification. This comprehensive dataset forms the foundation for the proposed two-stage fault classification framework, enabling the integration of data mining techniques to distinguish between normal operation, internal faults, and cyber-induced anomalies.

\section{Mathematical Modeling and Evaluation Methods of Proposed Framework} 
\begin{figure}[!t]
\centering
\includegraphics[width=\columnwidth,keepaspectratio]{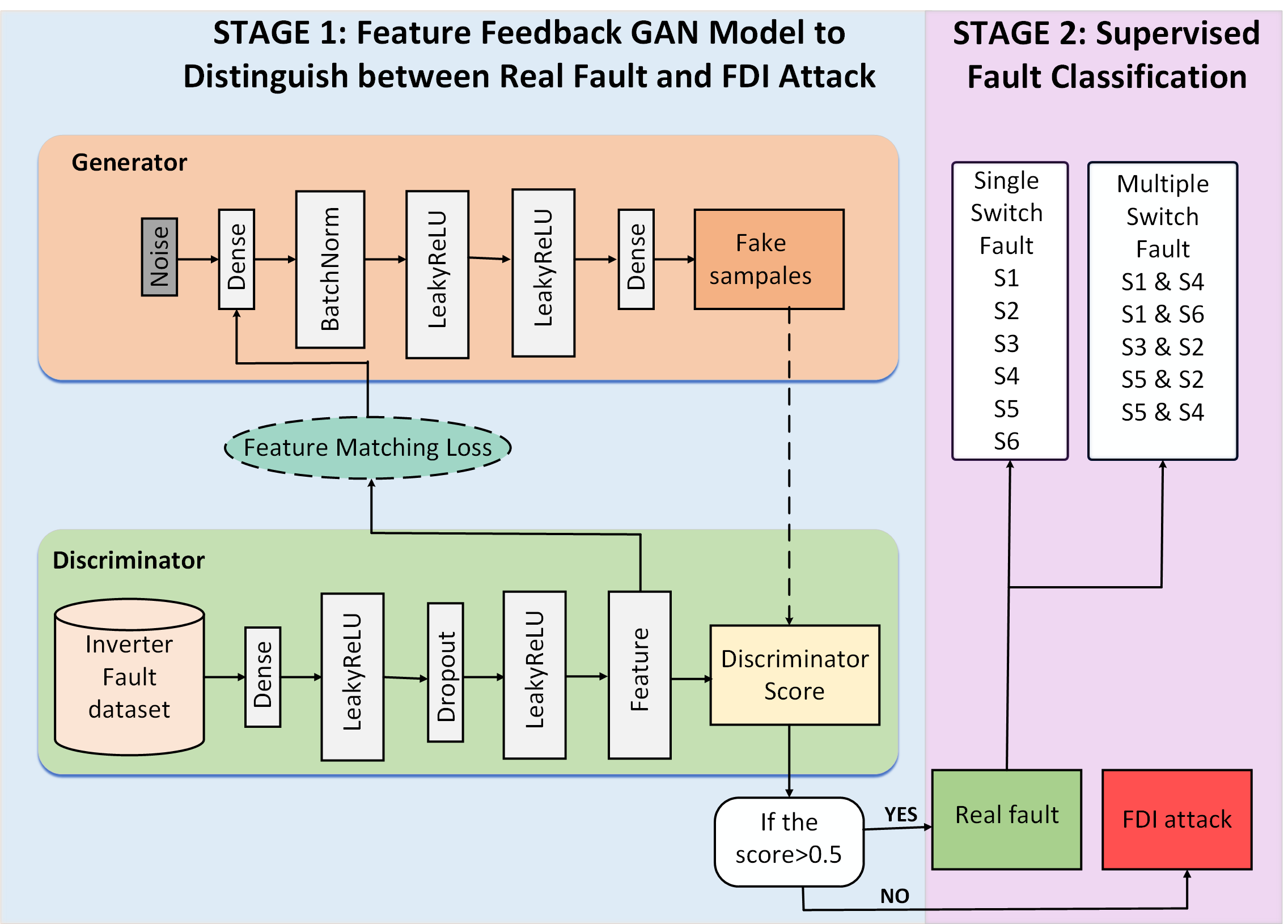}
\caption{The F2GAN architecture distinguishes between real inverter faults and FDI attacks and is coupled with supervised learning to classify and localize the real faults into specific single or multiple switch fault categories.}
\label{twostage}
\end{figure}

Fig. \ref{twostage} illustrates the architecture of the proposed F2GAN model. GANs are a powerful class of fully unsupervised models capable of learning to generate synthetic data that closely mimics the true data distribution. They have been successfully applied across various data types, including images, tabular data, and time-series signals. A conventional GAN operates as a two-player minimax game between two neural networks: a generator (G) and a discriminator (D). The generator attempts to produce synthetic samples that resemble real data, while the discriminator seeks to distinguish between real and generated samples. Both networks are trained simultaneously in a competitive manner, striving toward a Nash equilibrium where neither can improve without impacting the other.

However, traditional GANs primarily focus on binary classification (real vs. fake), often overlooking the intrinsic feature representations critical for system diagnosis. The proposed F2GAN framework incorporates a feature matching loss strategy to overcome this limitation. Rather than solely optimizing based on classification outcomes, F2GAN encourages the generator to minimize the difference between the intermediate feature activations (extracted from the discriminator) of real and fake samples. This guides the generator to synthesize data that appears realistic and preserves the underlying system characteristics.

Consequently, the discriminator learns a richer feature space that captures the normal operational patterns and fault signatures of inverter systems. Any deviation from these learned feature distributions—such as those caused by anomalies (e.g., FDI or zero-day attack) is effectively detected during testing. A threshold-based decision rule is applied: if the discriminator's output score exceeds 0.5, the input is classified as a real inverter fault; otherwise, it is flagged as a cyber-induced anomaly. The real fault samples are then passed to Stage 2, where supervised learning models further localize and classify the faults into specific single or multiple switch fault categories. The detailed mathematical modeling of the F2GAN framework is presented in the following section.
\subsection{Mathematical Modeling of F2GAN}

Let $x \sim rdata(x)$ represent the real data sampled from the true data distribution, and $z \sim p(z)$ denote the noise vector sampled from a prior distribution (commonly Gaussian or uniform). The generator maps the latent space $z$ to the data space, yielding $G(z)$. The discriminator outputs a scalar probability $D(x)$ representing the likelihood that the input $x$ is real. The classical GAN objective function is given by Eq. (\ref{eq:1})~\cite{10208147, 10318616, 8565906}: :
\begin{multline}
\label{eq:1}
\min_G \max_D V(D, G) = \mathbb{E}_{x \sim rdata}[\log D(x)] + \\
\mathbb{E}_{z \sim p(z)}[\log(1 - D(G(z)))]
\end{multline}

Here:
\begin{itemize}
    \item $D(x)$ outputs a probability:
    \[
    D(x) = 
    \begin{cases}
    1, & \text{if } x \text{ is real (from dataset)} \\
    0, & \text{if } x \text{ is fake (from generator)}
    \end{cases}
    \]
    \item $G(z)$ tries to fool the discriminator by generating data that $D$ classifies as real.
    \item The generator is trained to minimize $\log(1 - D(G(z)))$, also known as the \textbf{fooling loss}~\cite{10208147}.
\end{itemize}

In this application, the data includes scenarios affected by FDI attacks, which are unlabeled and cannot be handled using supervised learning. While a conventional GAN can effectively learn to generate realistic samples, its reliance solely on binary classification output from the discriminator (i.e., $D(x) = 1$ for real, $0$ for fake) limits its ability to capture fine-grained feature distribution differences. This is particularly problematic for subtle anomalies like FDI attacks that are statistically close to real data. 

Therefore, in F2GAN this limitation is addressed by incorporating a \textit{feature feedback mechanism}. Specifically, an intermediate feature representation is extracted from the discriminator, denoted as $f(x)$ for real samples and $f(G(z))$ for generated samples. These features are used to compute a secondary loss function, known as the \textbf{feature matching loss}, given by Eq. (\ref{eq:2}).

\begin{equation}
\label{eq:2}
\mathcal{L}_{\text{FM}} = \left\| \mathbb{E}_{x \sim rdata}[f(x)] - \mathbb{E}_{z \sim p(z)}[f(G(z))] \right\|_2^2
\end{equation}

The total generator loss in F2GAN thus becomes a combination of the fooling loss and feature feedback loss, as shown in Eq. (\ref{eq:3}):

\begin{equation}
\small
\label{eq:3}
\mathcal{L}_G = \underbrace{\mathbb{E}_{z \sim p(z)}\left[\log(1 - D(G(z)))\right]}_{\text{fooling loss}} + \lambda \cdot \underbrace{\left\| \mathbb{E}[f(x)] - \mathbb{E}[f(G(z))] \right\|_2^2}_{\text{feature feedback loss}}
\end{equation}

Here:
\begin{itemize}
    \item The first term penalizes the generator if $D(G(z))$ is low, encouraging it to generate more realistic samples.
    \item The second term ensures that generated samples closely resemble real samples in feature space.
    \item $\lambda$ is a hyperparameter to balance the importance of both terms.
\end{itemize}

The training progresses until the generator learns to produce samples whose features statistically match those of real data. As training approaches equilibrium, the discriminator’s outputs converge, with $D(x) \approx D(G(z)) \approx 0.5$, indicating that it can no longer confidently distinguish real from generated samples. Concurrently, the feature matching loss $\mathcal{L}_{\text{FM}}$ approaches zero, reflecting a high degree of similarity between the real and synthetic feature representations. Building upon this foundation, Algorithm~\ref{A2} outlines the complete integration of the F2GAN-based detection mechanism with the downstream fault classification pipeline. It summarizes the overall procedure, including the training dynamics, inference logic, and supervised fault classification based on labeled data.

\begin{figure}[h]
\centering
\begin{minipage}{\columnwidth}
\begin{algorithm}[H]
\caption{Two-Stage Framework: F2GAN-based Detection and Supervised Fault Classification}
\label{A2}
\begin{algorithmic}
\small
\State \textbf{Input:} Real fault dataset $x_i \in \mathbb{R}^d$, labels $y_i \in \mathcal{C}$
\State \textbf{Output:} Fault class prediction $\hat{y}$ or system diagnosis

\State \textbf{Stage 1: F2GAN Training with Real Fault Data}
    \State Initialize generator $G(z)$ and discriminator $D(x)$
    \State Extract intermediate feature layer $f(\cdot)$ from $D$
    \While{not converged}
        \State Sample real data batch $x \sim rdata$
        \State Sample noise $z \sim p(z)$ and generate $\hat{x} = G(z)$
        \State Compute discriminator loss:
        \[
        \mathcal{L}_D = -\mathbb{E}[\log D(x)] - \mathbb{E}[\log(1 - D(G(z)))]
        \]
        \State Compute generator fooling loss:
        \[
        \mathcal{L}_{G_\text{fool}} = \mathbb{E}[\log(1 - D(G(z)))]
        \]
        \State Compute feature matching loss:
        \[
        \mathcal{L}_{\text{FM}} = \left\| \mathbb{E}[f(x)] - \mathbb{E}[f(G(z))] \right\|_2^2
        \]
        \State Update generator with:
        \[
        \mathcal{L}_G = \mathcal{L}_{G_\text{fool}} + \lambda \cdot \mathcal{L}_{\text{FM}}
        \]
    \EndWhile

\State \textbf{Stage 1 Inference: FDI vs Internal Fault Detection}
    \State For test sample $x_{\text{test}}$, compute $D(x_{\text{test}})$
    \If{$D(x_{\text{test}}) > 0.5$}
        \State Sample is classified as Real Internal Fault
        \State Proceed to Stage 2 for fault classification
    \Else
        \State Sample is classified as FDI Attack (Anomaly)
        \State Terminate: No classification
    \EndIf

\State \textbf{Stage 2: Fault Classification using Supervised Learning}
    \State Train classifier $f: \mathbb{R}^d \rightarrow \mathcal{C}$ on $(x_i, y_i)$
    \State Predict class label for real fault:
    \[
    \hat{y} = f(x_{\text{test}})
    \]

\State \textbf{Output:} $\hat{y}$ for fault type or anomaly flag

\end{algorithmic}
\end{algorithm}
\end{minipage}
\end{figure}

\subsection{Modelling of FDI Attack}

FDI attacks are a class of cyber threats where adversaries strategically alter measurement data to mislead decision-making in power system operations. These attacks pose a serious threat in smart grid environments where measurements are used to estimate the system state, and control decisions are made based on these estimations.

Let the linearized measurement model be expressed as in Eq. (\ref{eq4})~\cite{8731755}:
\begin{equation}
\label{eq4}
z = Hx + e
\end{equation}
where:
\begin{itemize}
    \item $z \in \mathbb{R}^m$ is the measurement vector,
    \item $H \in \mathbb{R}^{m \times n}$ is the measurement matrix (Jacobian),
    \item $x \in \mathbb{R}^n$ is the true system state vector,
    \item $e \in \mathbb{R}^m$ is the Gaussian measurement noise vector.
\end{itemize}

In a typical state estimation process, the estimate $\hat{x}$ is obtained by minimizing the residual r is given by the Eq.(\ref{5}).
\begin{equation}
\label{5}
r = \| z - H\hat{x} \|
\end{equation}

An FDI attack modifies the original measurement vector by injecting a malicious vector $a \in \mathbb{R}^m$, yielding a corrupted measurement $z_{a}$ given by Eq. (\ref{6}).
\begin{equation}
\label{6}
z_a = z + a= Hx + e + a
\end{equation}

If the attacker has knowledge of the system model $H$, the attack vector can be designed as in Eq. (\ref{7}).
\begin{equation}
\label{7}
a = Hc
\end{equation}
for some arbitrary vector $c \in \mathbb{R}^n$. The resulting attacked measurements are expressed as in Eqs. (\ref{8}), (\ref{9}).
\begin{equation}
\label{8}
z_a = Hx + e + Hc = H(x + c) + e
\end{equation}

This implies that:
\begin{equation}
\label{9}
\hat{x}_a = x + c
\end{equation}
and the residual remains as in Eq. (\ref{Eq:14})~\cite{10793094}:
\begin{equation}
\label{Eq:14}
r_a = \| z_a - H\hat{x}_a \| = \| H(x + c) + e - H(x + c) \| = \| e \|
\end{equation}

Since the residual $r_a$ is unaffected and statistically identical to the original noise vector $e$, such an attack is \textbf{unobservable} to conventional bad data detection methods. This makes FDI attacks extremely dangerous, as they can manipulate system operations without triggering alarms.

In this work, these attacks are simulated by injecting perturbations into voltage and current measurements within the communication layer of the microgrid model, mimicking real-world unobservable FDI scenarios. The proposed F2GAN-based model is then evaluated to distinguish between real internal faults and stealthy FDI anomalies.

\subsection{Mathematical Modeling of Fault Classification}

Once the F2GAN discriminator has verified that a given data instance corresponds to a real internal fault (as opposed to an FDI attack), the fault localization and classification stage is initiated using supervised learning algorithms.

The dataset used consists of 1,097 labeled instances, each with 16 signal features including voltage, current, frequency, and their transformed components. The fault type is encoded in a categorical label identifying specific inverter switch faults.

Let the dataset be represented as in Eq.~(\ref{11}):
\begin{equation}
\label{11}
\mathcal{D} = \{(\mathbf{s}_u, \omega_u)\}_{u=1}^{T}, \quad T = 1097
\end{equation}
where:
\begin{itemize}
    \item $\mathbf{s}_u \in \mathbb{R}^{16}$ is the $u$-th signal sample (feature vector),
    \item $\omega_u \in \Omega$ is the corresponding class label for that sample,
    \item $\Omega$ is the set of predefined fault classes.
\end{itemize}

The goal is to learn a classifier $g: \mathbb{R}^{16} \rightarrow \Omega$ such that the predicted class $\hat{\omega}_u = g(\mathbf{s}_u)$ is as close as possible to the true class $\omega_u$.

\subsubsection{Support Vector Machine (SVM)}
SVM finds the optimal hyperplane that separates fault classes by maximizing the margin. The mathematical equation is given by Eq. (\ref{12})~\cite{9625829}:
\begin{equation}
\label{12}
g_{\text{SVM}}(\mathbf{s}) = \text{sign}(\mathbf{w}^\top \mathbf{s} + b)
\end{equation}
where $\mathbf{w}$ is the weight vector and $b$ is the bias term.

\subsubsection{K-Nearest Neighbors (KNN)}
KNN classifies a new input by majority voting among its $k$ nearest neighbors. Mathematically it is represented as in Eq. (\ref{13})~\cite{9777035}:
\begin{equation}
\label{13}
g_{\text{KNN}}(\mathbf{s}) = \arg\max_{\omega \in \Omega} \sum_{j \in \mathcal{N}_k(\mathbf{s})} \mathbb{I}[\omega_j = \omega]
\end{equation}
where $\mathcal{N}_k(\mathbf{s})$ denotes the $k$ nearest neighbors and $\mathbb{I}[\cdot]$ is the indicator function.

\subsubsection{Decision Tree (DT)}
Decision Trees classify inputs by recursively navigating a tree structure based on feature thresholds, where each branch represents a decision criterion aiding in fault classification. The impurity of a node is quantified using the Gini index, as shown in Eq. (\ref{14}):
\begin{equation}
\label{14}
G(T) = \sum_{i=1}^{n} p_i \times \left( 1 - p_i \right)
\end{equation}
where \( G(T) \) is the Gini impurity of node \( T \), and \( p_i \) represents the probability of a sample being classified as class \( i \). 

\subsubsection{Artificial Neural Network (ANN)}
A feedforward neural network maps input vectors to output classes through layered transformations, as expressed in Eq. (\ref{15}):
\begin{equation}
\label{15}
g_{\text{ANN}}(\mathbf{s}) = \sigma(\mathbf{W}_2 \cdot \phi(\mathbf{W}_1 \mathbf{s} + \mathbf{b}_1) + \mathbf{b}_2)
\end{equation}
where:
\begin{itemize}
    \item $\mathbf{W}_1, \mathbf{W}_2$ are weight matrices,
    \item $\mathbf{b}_1, \mathbf{b}_2$ are bias vectors,
    \item $\phi(\cdot)$ denotes a non-linear activation function (e.g., ReLU),
    \item $\sigma(\cdot)$ denotes the output activation function (e.g., softmax).
\end{itemize}

These models are trained using the labeled internal fault dataset and evaluated using metrics such as accuracy, precision, recall, and F1-score to ensure robust classification performance.
\vspace{-1.5mm}
\section{Evaluation and results}
The proposed microgrid framework was developed in the MATLAB/Simulink environment, incorporating three distributed renewable energy (DRE) sources and a centralized control system. This setup generated an internal fault dataset by introducing systematic fault scenarios at each inverter switch. The dataset comprises six single-switch fault cases and six multiple-switch fault combinations, simulated under various operating conditions, including load fluctuations, power level variations, and mode transitions (e.g., grid-connected to islanded). A total of 1,097 cases were generated, each containing 16 extracted feature variables and categorized into 12 distinct fault classes.

This dataset was used to implement and evaluate a conventional GAN (CGAN) and the proposed Feature-Matching GAN (F2GAN) in Python. The hyperparameters used in both models are detailed in Table \ref{tab:hyperparam_comparison}. 
\begin{table}[htbp]
\caption{\centering Hyperparameter Comparison: Conventional GAN vs. Proposed F2GAN}
\label{tab:hyperparam_comparison}
\centering
\resizebox{\columnwidth}{!}{%
\begin{tabular}{|l|c|c|}
\hline
\textbf{Parameter} & \textbf{Conventional GAN} & \textbf{F2GAN (Proposed)} \\
\hline
Input Features Dimension & 16 & 16 \\
\hline
Latent Dimension & 32 & 64 \\
\hline
Generator Layers & 3 (64-128-Out) & 4 (256-512-1024-Out) \\
\hline
Generator Activations & ReLU + Tanh & LeakyReLU + Tanh \\
\hline
Discriminator Layers & 3 (128-64-1) & 4 (1024-512-256-128) \\
\hline
Discriminator Activations & ReLU + Sigmoid & LeakyReLU + Sigmoid \\
\hline
Dropout in Discriminator & None & 0.3 (applied to 2 layers) \\
\hline
Feature Matching Loss & Not Used & \textbf{Used} \\
\hline
Epochs & 5000 & 5000 \\
\hline
Batch Size & 64 & 64 \\
\hline
\end{tabular}%
}
\end{table}
80$\%$  of the dataset was utilized for training the discriminator. For testing, a separate dataset was prepared by combining synthetic FDI attack data with the remaining 20$\%$ of the internal fault data.

\begin{figure}[htbp]
  \centering

  \begin{subfigure}[t]{0.48\columnwidth}
    \centering
    \includegraphics[width=\linewidth]{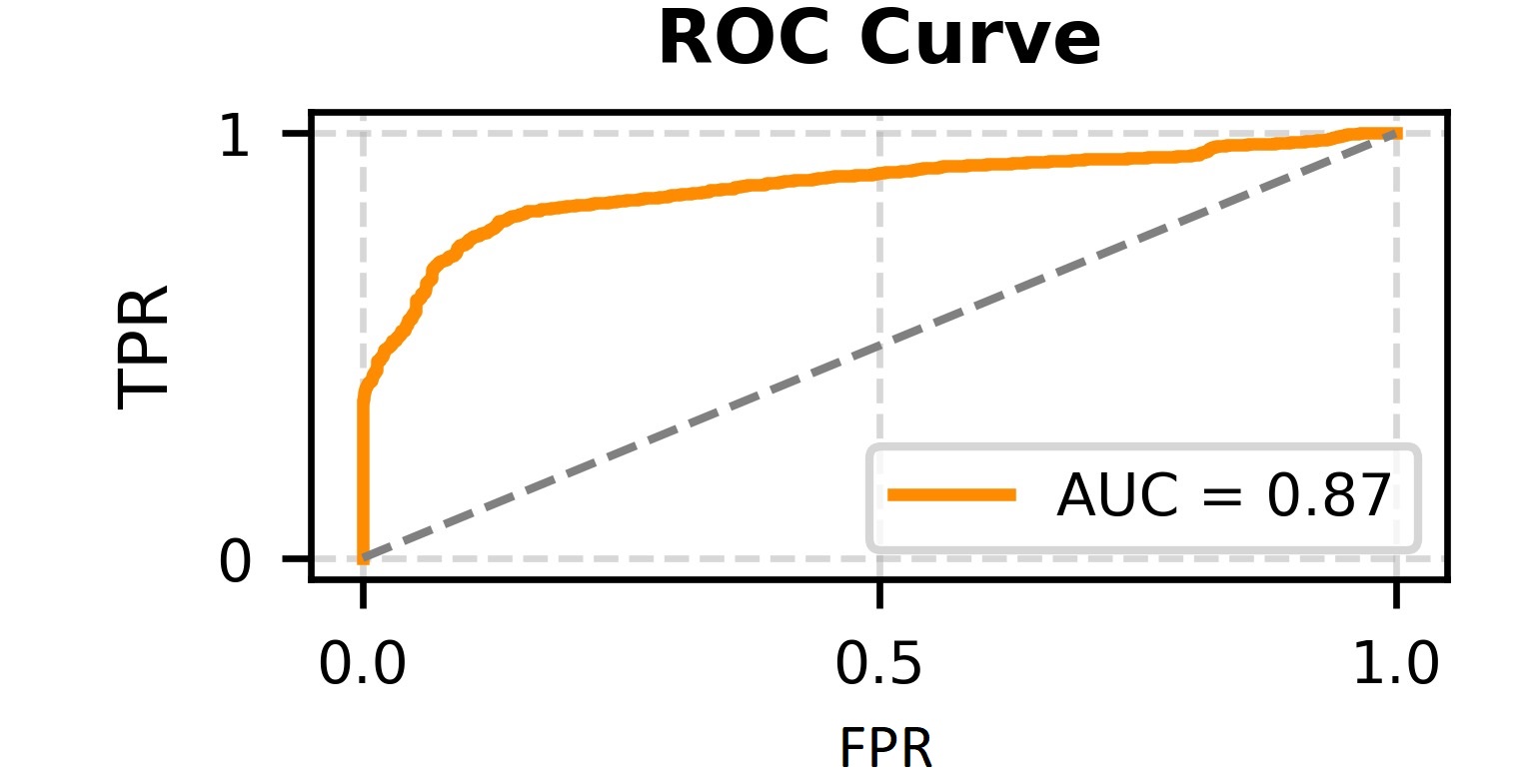}
    \caption{F2GAN Discriminator (AUC = 0.87)}
    \label{fig:roc_f2gan}
  \end{subfigure}
  \hfill
  \begin{subfigure}[t]{0.48\columnwidth}
    \centering
    \includegraphics[width=\linewidth]{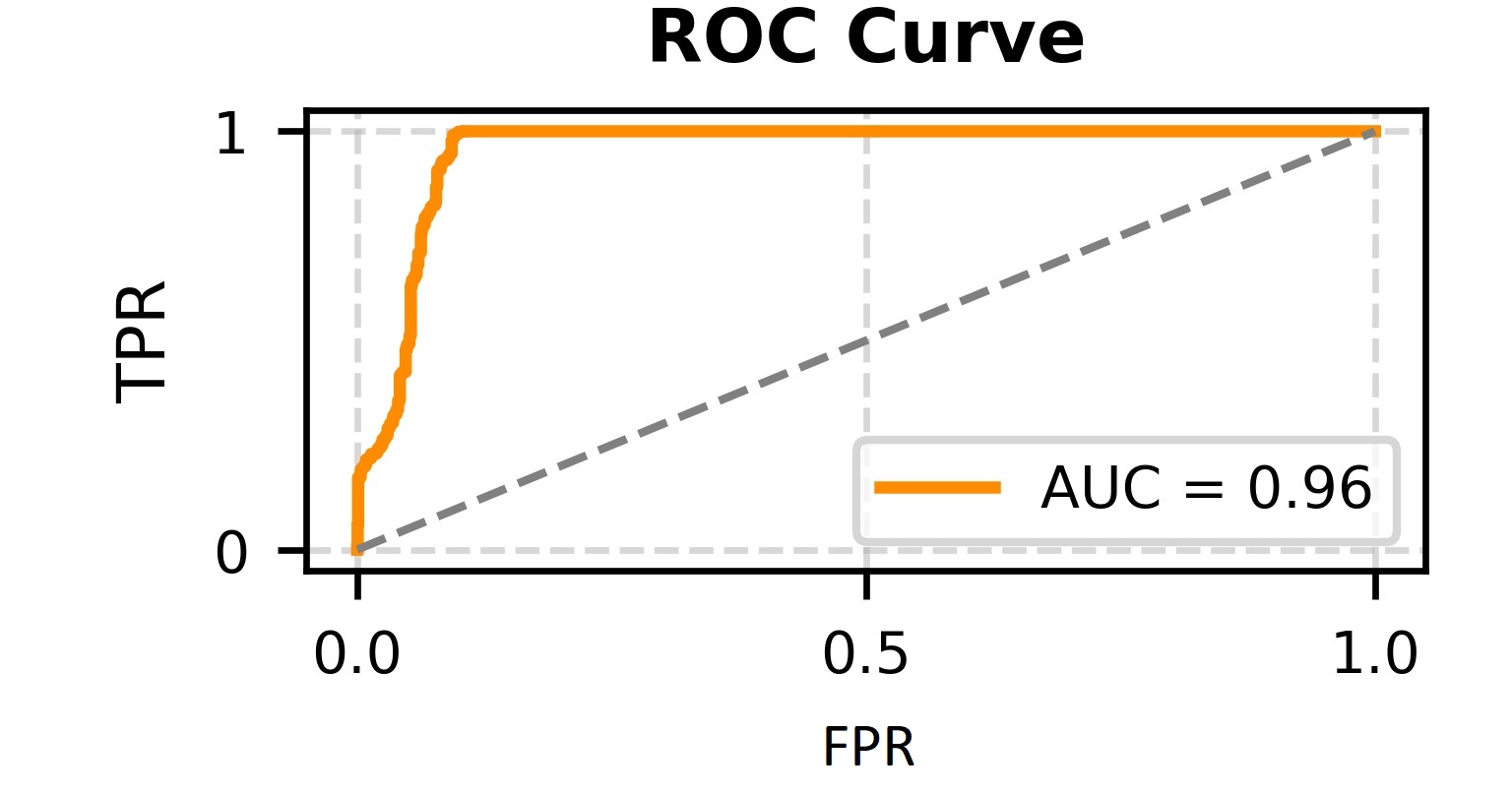}
    \caption{F2GAN Discriminator (AUC = 0.96))}

  \end{subfigure}
  \vspace{0.4cm}
  \begin{subfigure}[t]{0.48\columnwidth}
    \centering
    \includegraphics[width=\linewidth]{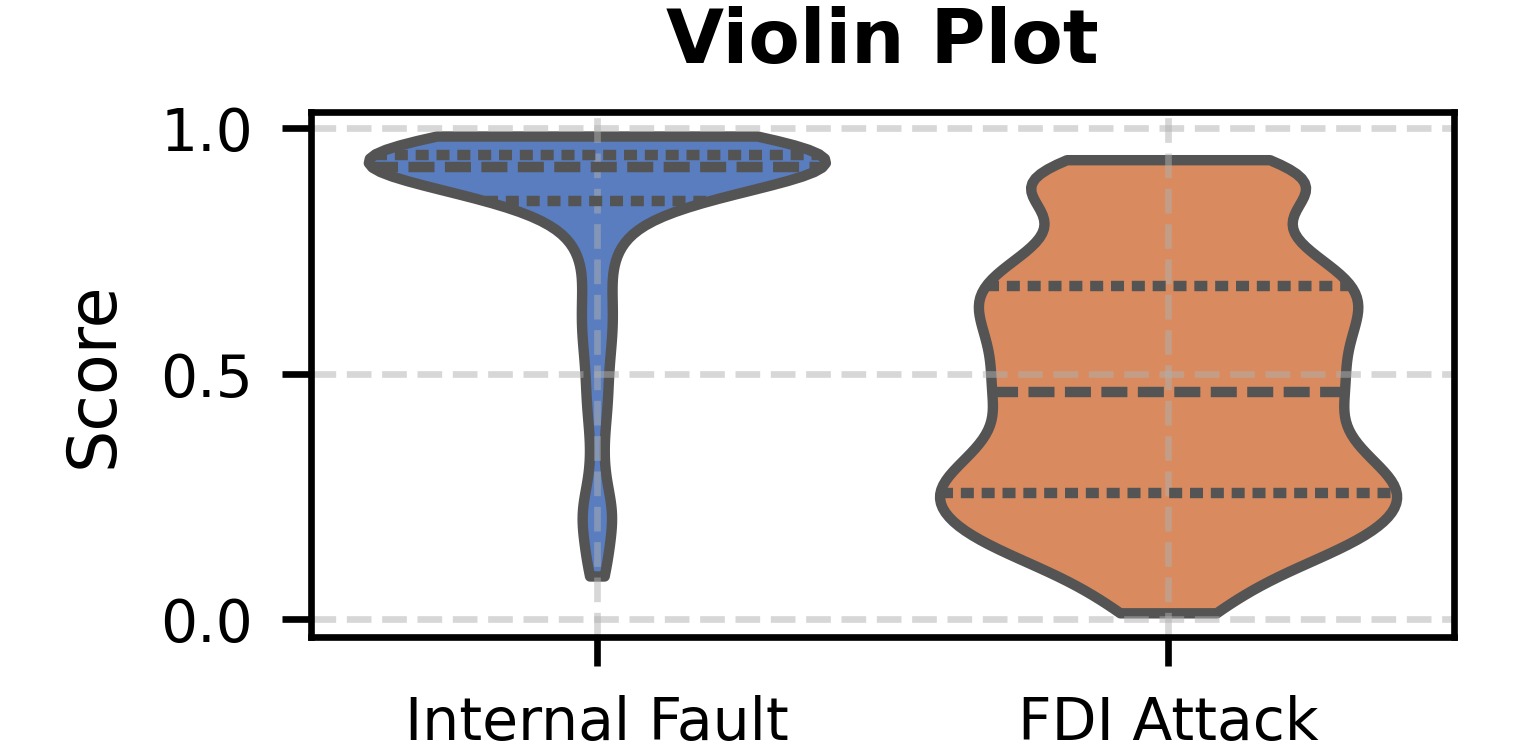}
    \caption{Violin plot of feature distributions without feedback}
    \label{fig:violin}
  \end{subfigure}
  \hfill
  \begin{subfigure}[t]{0.48\columnwidth}
    \centering
    \includegraphics[width=\linewidth]{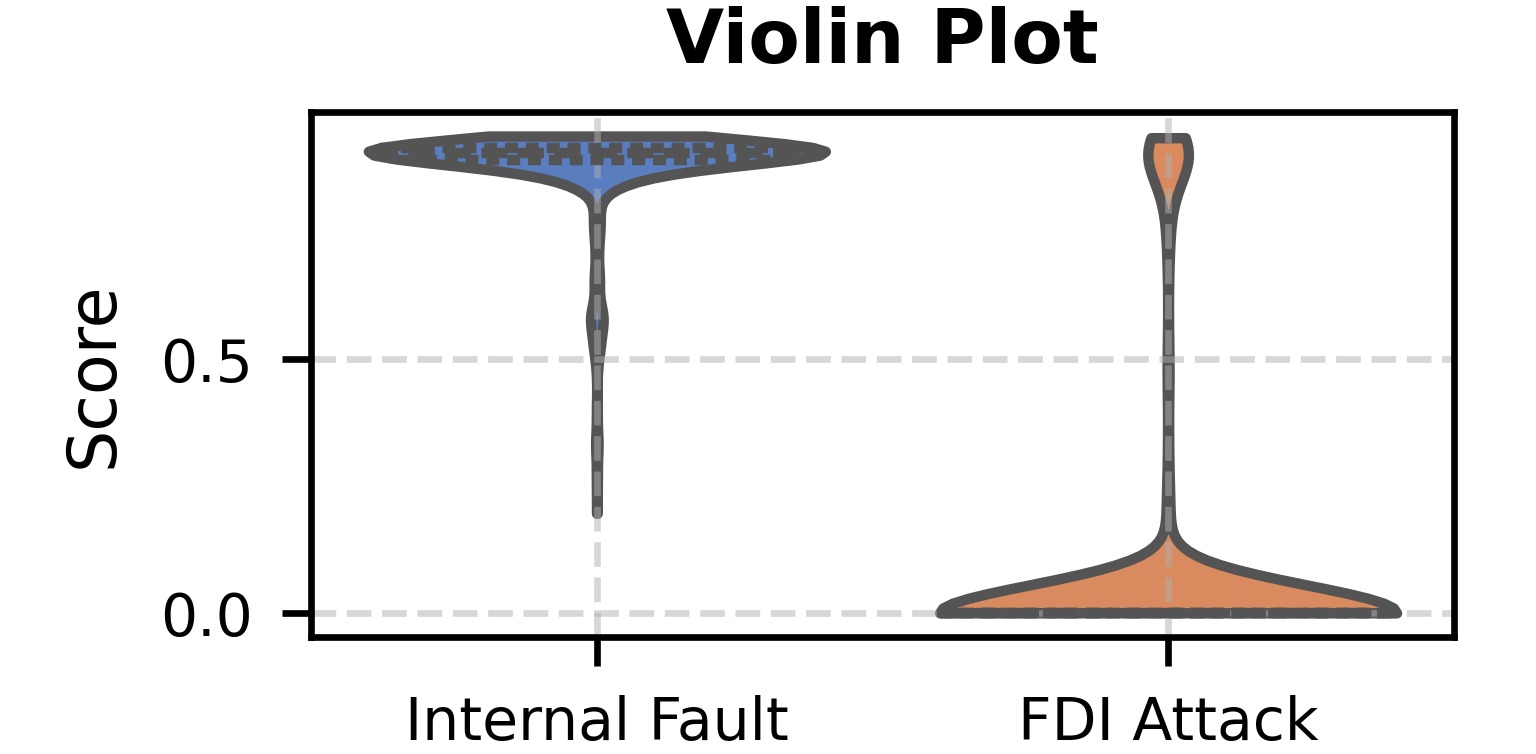}
    \caption{Violin plot of feature distributions with feedback}
   
  \end{subfigure}
  \caption{Performance comparison and system-level visualization across models and data layers.}
  \label{fig:results_compact}
\end{figure}


Fig.~\ref{fig:results_compact} illustrates the comparative performance between the proposed F2GAN and the conventional GAN models. Subfigures~\ref{fig:results_compact}(a) and~\ref{fig:results_compact}(b) present the ROC curves, where F2GAN achieves a higher AUC of 0.96, demonstrating superior capability in distinguishing internal faults from FDI attacks, compared to the baseline GAN with an AUC of 0.87. Subfigures~\ref{fig:results_compact}(c) and \ref{fig:results_compact}(d) depict the violin plots of discriminator score distributions, clearly showing that F2GAN assigns well-separated score ranges to real faults and cyber anomalies. This separation indicates more confident and less ambiguous classification by F2GAN, effectively reducing confusion between fault types and attacks.

Table \ref{tab:gan_f2gan_comparison} illustrates the statistical evaluation of the proposed F2GAN architecture which demonstrates a clear performance advantage over the conventional GAN in distinguishing internal faults from FDI attacks. F2GAN achieves a significantly higher accuracy (93.27\% vs. 65.52\%), along with improved precision (82.89\%), recall (98.91\%), and F1-score (90.19\%), indicating a more reliable and balanced fault detection. The AUC also improves from 0.8747 to 0.9562, reflecting superior classification confidence across thresholds. Score distributions reveal that F2GAN assigns high, consistent scores to internal faults (mean = 0.8797, std = 0.0907) and low scores to FDI attacks (mean = 0.0859), unlike the conventional GAN which shows greater confusion. Furthermore, the lower KL divergence (7.0165 vs. 8.5098) confirms F2GAN’s stronger separation of real and anomalous data. These findings validate the feature feedback mechanism’s effectiveness in improving the generator's realism and enhancing the discriminator’s sensitivity to subtle or zero-day cyber anomalies.

\begin{table}[htbp]
\caption{\centering Statistical Evaluation Metrics for Conventional GAN and F2GAN}
\label{tab:gan_f2gan_comparison}
\centering
\resizebox{\columnwidth}{!}{%
\begin{tabular}{|c|c|c|}
\hline
\textbf{Metric} & \textbf{Conventional GAN} & \textbf{F2GAN (Proposed)} \\
\hline
Accuracy & 0.6552 & 0.9327 \\
\hline
Precision & 0.4729 & 0.8289 \\
\hline
Recall & 0.8979 & 0.9891 \\
\hline
F1 Score & 0.6195 & 0.9019 \\
\hline
AUC (ROC) & 0.8747 & 0.9562 \\
\hline
\textbf{Discriminator Score Distribution} & & \\
\hline
Mean (Inverter Faults) & 0.8322 & 0.8797 \\
\hline
Std Dev (Inverter Faults) & 0.2131 & 0.0907 \\
\hline
Mean (FDI attack) & 0.4757 & 0.0859 \\
\hline
Std Dev (FDI attack) & 0.2537 & 0.2578 \\
\hline
\textbf{KL Divergence (Real vs. FDI)} & 8.5098 & 7.0165 \\
\hline
\end{tabular}%
}
\end{table}
Table~\ref{S4} shows the performance of four classifiers in detecting inverter faults. The ANN model outperformed all others with an accuracy, precision, recall, and F1-score of 99.99\%. The DT and KNN models also performed well, with DT achieving 97.87\% accuracy and KNN reaching a recall of 97.60\%. The SVM model had the lowest performance among the four, with an accuracy of 95.74\% and an F1-score of 95.99\%.

\begin{table}[h]
\centering
\caption{\centering Performance Metrics for inverter fault classifications}
\label{S4}
\scriptsize
\begin{tabular}{|c|c|c|c|c|}
\hline
\textbf{Model} & \textbf{Accuracy (\%)} & \textbf{Precision (\%)} & \textbf{Recall (\%)} & \textbf{F1-Score (\%)} \\ \hline
DT  & 97.87 & 97.44 & 97.35 & 97.35 \\ \hline
KNN & 97.16 & 97.43 & 97.60 & 97.48 \\ \hline
SVM & 95.74 & 96.08 & 96.19 & 95.99 \\ \hline
ANN & 99.99 & 99.99 & 99.99 & 99.99 \\ \hline
\end{tabular}
\end{table}

\section{Conclusion and future scope}
This study proposes a robust two-stage framework that integrates an unsupervised F2GAN model for system diagnosis with a supervised learning system for fault classification in inverter-based microgrids. The approach effectively differentiates real internal faults from cyber-induced FDI anomalies, achieving high accuracy and outperforming conventional GAN-based methods. The results highlight the model’s capability to generalize well, even in the presence of subtle or zero-day attacks. As part of future work, the framework can be further improved by incorporating synthetic data generation techniques to enrich fault scenarios and enhance the classifier's robustness. While the current focus is on internal faults, the methodology can be extended to detect and classify external faults within distribution systems. Moreover, beyond system diagnosis, the use of GAN-based and large language model (LLM)-assisted recovery mechanisms can be explored to reconstruct the true system state from compromised data, facilitating real-time correction and resilient microgrid control.

\bibliographystyle{IEEEtran}
\bibliography{IEEEabrv,ref}

\begin{thebibliography}{10}
\providecommand{\url}[1]{#1}
\csname url@samestyle\endcsname
\providecommand{\newblock}{\relax}
\providecommand{\bibinfo}[2]{#2}
\providecommand{\BIBentrySTDinterwordspacing}{\spaceskip=0pt\relax}
\providecommand{\BIBentryALTinterwordstretchfactor}{4}
\providecommand{\BIBentryALTinterwordspacing}{\spaceskip=\fontdimen2\font plus
\BIBentryALTinterwordstretchfactor\fontdimen3\font minus \fontdimen4\font\relax}
\providecommand{\BIBforeignlanguage}[2]{{%
\expandafter\ifx\csname l@#1\endcsname\relax
\typeout{** WARNING: IEEEtran.bst: No hyphenation pattern has been}%
\typeout{** loaded for the language `#1'. Using the pattern for}%
\typeout{** the default language instead.}%
\else
\language=\csname l@#1\endcsname
\fi
#2}}
\providecommand{\BIBdecl}{\relax}
\BIBdecl

\bibitem{8590418}
Y.~Mei and H.~Yuan, ``A novel open-circuit fault diagnosis method for voltage source inverter,'' in \emph{2018 IEEE International Power Electronics and Application Conference and Exposition (PEAC)}, 2018, pp. 1--6.

\bibitem{8329516}
Z.~Huang, Z.~Wang, and H.~Zhang, ``A diagnosis algorithm for multiple open-circuited faults of microgrid inverters based on main fault component analysis,'' \emph{IEEE Transactions on Energy Conversion}, vol.~33, no.~3, pp. 925--937, 2018.

\bibitem{9544141}
M.~Azimipanah, M.~Hassanifar, and Y.~Neyshabouri, ``Open circuit fault detection and diagnosis for seven-level hybrid active neutral point clamped (anpc) multilevel inverter,'' in \emph{2021 29th Iranian Conference on Electrical Engineering (ICEE)}, 2021, pp. 268--273.

\bibitem{10214285}
C.~N. Ibem, M.~E. Farrag, A.~A. Aboushady, and S.~M. Dabour, ``Multiple open switch fault diagnosis of three phase voltage source inverter using ensemble bagged tree machine learning technique,'' \emph{IEEE Access}, vol.~11, pp. 85\,865--85\,877, 2023.

\bibitem{10215460}
R.~Peykarporsan, J.~Heidary, S.~Oshnoei, and T.~T. Lie, ``A machine learning approach for fault detection and diagnosis in four-legged inverters,'' in \emph{2023 IEEE Kansas Power and Energy Conference (KPEC)}, 2023, pp. 1--5.

\bibitem{10451077}
Z.~Wu and J.~Zhao, ``Open-circuit fault diagnosis for grid-tied t-type inverters using an lstm autoencoder,'' in \emph{2023 China Automation Congress (CAC)}, 2023, pp. 3334--3339.

\bibitem{9970321}
M.~Baker, A.~Y. Fard, H.~Althuwaini, and M.~B. Shadmand, ``Real-time ai-based anomaly detection and classification in power electronics dominated grids,'' \emph{IEEE Journal of Emerging and Selected Topics in Industrial Electronics}, vol.~4, no.~2, pp. 549--559, 2023.

\bibitem{8731755}
Y.~Li, Y.~Wang, and S.~Hu, ``Online generative adversary network based measurement recovery in false data injection attacks: A cyber-physical approach,'' \emph{IEEE Transactions on Industrial Informatics}, vol.~16, no.~3, pp. 2031--2043, 2020.

\bibitem{10689021}
L.~Garza, P.~Mandal, and G.~Ravikumar, ``Mitigation of cyber attacks in der-integrated distribution grid using deep generative models,'' in \emph{2024 IEEE Power \& Energy Society General Meeting (PESGM)}, 2024, pp. 1--5.

\bibitem{10208147}
Y.~Hu, Y.~Li, L.~Song, H.~P. Lee, P.~J. Rehm, M.~Makdad, E.~Miller, and N.~Lu, ``Multiload-gan: A gan-based synthetic load group generation method considering spatial-temporal correlations,'' \emph{IEEE Transactions on Smart Grid}, vol.~15, no.~2, pp. 2309--2320, 2024.

\bibitem{10318616}
D.~Kim, S.~Majumder, and L.~Xie, ``Line-post insulator fault classification model using deep convolutional gan-based synthetic images,'' in \emph{2023 North American Power Symposium (NAPS)}, 2023, pp. 1--6.

\bibitem{8565906}
J.~Liu, F.~Qu, X.~Hong, and H.~Zhang, ``A small-sample wind turbine fault detection method with synthetic fault data using generative adversarial nets,'' \emph{IEEE Transactions on Industrial Informatics}, vol.~15, no.~7, pp. 3877--3888, 2019.

\bibitem{10793094}
K.~Miao, M.~Zhang, F.~Guo, R.~Lu, and X.~Guan, ``Detection of false data injection attacks in smart grids: An optimal transport-based reliable self-training approach,'' \emph{IEEE Transactions on Information Forensics and Security}, vol.~20, pp. 709--723, 2025.

\bibitem{9625829}
A.~Srivastava and S.~Parida, ``A robust fault detection and location prediction module using support vector machine and gaussian process regression for ac microgrid,'' \emph{IEEE Transactions on Industry Applications}, vol.~58, no.~1, pp. 930--939, 2022.

\bibitem{9777035}
A.~Kumar, E.~Koley, and A.~G. Rameshrao, ``An intelligent fault detection and faulty line identification scheme for hybrid microgrid using ensemble of knn approach,'' in \emph{2022 Second International Conference on Power, Control and Computing Technologies (ICPC2T)}, 2022, pp. 1--6.

\end{thebibliography}

\end{document}